\documentclass[runningheads]{llncs}
\usepackage[T1]{fontenc}
\usepackage{graphicx}
\usepackage[ruled,vlined]{algorithm2e}
\usepackage{amsmath}
\usepackage{graphicx}  
\usepackage{subfigure}    
\usepackage{float}     
\usepackage{booktabs} 
\usepackage{multirow}
\usepackage{marvosym}

\usepackage{tikz}
\usepackage{etoolbox}
\usepackage{enumitem}
\newcommand*{\circled}[1]{\lower.7ex\hbox{\tikz\draw (0pt, 0pt)%
    circle (.5em) node {\makebox[1em][c]{\small #1}};}}
\robustify{\circled}

\begin{document}
\title{FedDCT: A Dynamic Cross-Tier Federated Learning Framework in Wireless Networks}
%
%
\author{Youquan Xian\inst{1,2} \and
Xiaoyun Gan\inst{1,2} \and
Chuanjian Yao\inst{1,2} \and
Dongcheng Li\inst{1,2} \and
Peng Wang\inst{1,2} \and
Peng Liu\inst{1,2}\textsuperscript{(\Letter)} \and
Ying Zhao\inst{3}\textsuperscript{(\Letter)}
}
\authorrunning{Xian et al.}
%
\institute{Key Lab of Education Blockchain and Intelligent Technology, Ministry of Education, Guangxi Normal University, Guilin 54104, China 
\and
School of Computer Science and Engineering, Guangxi Normal University, Guilin 54104, China
\and
School of Business, Guilin University of Electronic Technology, Guilin 54104, China
\email{xianyouquan@stu.gxnu.edu.cn,liupeng@gxnu.edu.cn,zhaoying@guet.edu.cn}}
\maketitle              
\begin{abstract}
Federated Learning (FL), as a privacy-preserving machine learning paradigm, trains a global model across devices without exposing local data. However, resource heterogeneity and inevitable stragglers in wireless networks severely impact the efficiency and accuracy of FL training. In this paper, we propose a novel Dynamic Cross-Tier Federated Learning framework (FedDCT). Firstly, we design a dynamic tiering strategy that dynamically partitions devices into different tiers based on their response times and assigns specific timeout thresholds to each tier to reduce single-round training time. Then, we propose a cross-tier device selection algorithm that selects devices that respond quickly and are conducive to model convergence to improve convergence efficiency and accuracy. Experimental results demonstrate that the proposed approach under wireless networks outperforms the baseline approach, with an average reduction of 54.7\% in convergence time and an average improvement of 1.83\% in convergence accuracy.

\keywords{Wireless networks  \and Federated learning \and Resource heterogeneity.}
\end{abstract}

\section{Introduction}
Driven by the rapid growth of distributed data mining, Federated Learning (FL) has garnered significant attention from both the academic and industrial sectors due to its nature of distributed training and privacy preservation \cite{duan2023combining}. FL enables the training of a global model across devices without exposing local data. The FL process can be summarized as follows: the server initializes the global model and selects devices to distribute the global model. The chosen devices train using the obtained global model and local data, and the trained models are then uploaded to the server. Finally, the server applies aggregation algorithms such as weighted averaging (e.g., FedAvg \cite{mcmahan2017communication}) to aggregate the uploaded models into the global model, and subsequently selects new participating devices to distribute the aggregated new model.

In wireless networks, devices often exhibit heterogeneity in computational and communication resources, and issues such as communication failures or device malfunctions can result in a significant number of devices dropping out. Dropout devices or those with lower computational capabilities may lag significantly behind other devices, leading to inefficiencies in a single round of FL training \cite{wang2022asynchronous,ye2023heterogeneous}.
To mitigate the adverse effects of resource heterogeneity on FL, FedMCCS \cite{abdulrahman2020fedmccs} predicts whether devices can complete tasks based on their computational resources and communication capabilities, maximizing the selection of devices to enhance convergence speed. Leng et al. \cite{leng2022client} and Zhang et al. \cite{zhang2023enhancing}, from the perspective of network resources, allocated sufficient network resources to training devices to reduce training time. Similarly, Zhang et al. \cite{zhang2022joint} employed reinforcement learning to select participating devices and allocate different local iteration numbers and network resources to participants.
However, device dropout in practical networks is unavoidable. While asynchronous FL no longer requires waiting for other devices to upload model parameters in each training round, avoiding dropout issues \cite{xie2019asynchronous}, asynchronous FL typically accompanies the model staleness effect, leading to difficulties in model convergence \cite{liu2024fedasmu}. Moreover, the aforementioned approaches, aiming for efficiency, overly focus on resource-rich devices, exacerbating the disparity in training participation among devices, causing model drift, and reducing model convergence accuracy \cite{huang2020efficiency}.

Thus, TiFL \cite{chai2020tifl} proposed the concept of tiered FL, dividing devices into different tiers based on their training response times, and then randomly selecting devices from each tier to participate in training. It not only reduces the disparity in single-round device training times, improving single-round training efficiency but also conducts training on a tier-by-tier basis, alleviating the impact of model drift \cite{pfeiffer2023federated}. However, tiered FL methods like TiFL still face challenges of underutilized resources, and their simplistic tiering approach fails to accurately partition devices, especially in cases of resource heterogeneity and device dropout. Therefore, the central issue is how to dynamically partition devices in a wireless network environment while improving training efficiency without causing model drift.

While asynchronous FL \cite{xie2019asynchronous} significantly boosts the efficiency of a single round of training by eliminating the need to wait for lagging devices, asynchronous FL training often requires more iterations and incurs higher communication overhead \cite{xu2021asynchronous,chai2021fedat}. Additionally, it is difficult to combine with the existing synchronous FL applications \cite{chai2020tifl}. Therefore, TiFL \cite{chai2020tifl} introduces the concept of tiered FL, categorizing devices into different tiers based on their training response times. Devices are then randomly selected from each tier to participate in training, reducing the disparity in individual device training times and enhancing the efficiency of a single round of training. However, tiered FL solutions like TiFL only address the reduction of resource heterogeneity among devices in a single training round and do not consider the possibility of devices dropping out in wireless networks, potentially leading to a significant increase in the waiting time for a single round. Therefore, a central challenge remains: how to improve the convergence efficiency and accuracy of FL in the presence of resource heterogeneity and dropout issues in wireless networks.

In this paper, we propose a novel Dynamic Cross-Tier Federated Learning framework (FedDCT), aiming to maximize convergence efficiency while avoiding model drift. This framework comprises two core modules: the dynamic tiering module and the cross-tier client selection module, which can be seamlessly integrated with existing FL applications in a non-intrusive manner. Firstly, the dynamic tiering module dynamically evaluates the response times of clients and categorizes them into different logical tiers, assigning specific timeout thresholds to each tier. Then, the cross-tier client selection module selects devices for FL training that exhibit fast response times and facilitate model convergence. The main contributions of this paper are as follows:

\begin{itemize}
\item To address the challenges of resource heterogeneity and device dropout in wireless networks, we design a dynamic tiering strategy. It involves real-time evaluation of device response times, tiering, and assigning specific timeout thresholds to each tier, enhancing the convergence efficiency of FL.

\item We propose a cross-tier client selection strategy. It first adaptively selects tiers that facilitate model convergence and exhibit fast response times, as well as the participating devices within those tiers. Effectively optimizing the utilization of idle resources, enhancing convergence speed and accuracy.

\item Through simulation experiments, we verify that the proposed approach in wireless network scenarios, compared to the baseline solution, achieves an average reduction of 54.7\% in convergence time and an average improvement of 1.83\% in convergence accuracy.
\end{itemize}

\section{FedDCT: Dynamic Cross-Tier Federated Learning}

\begin{figure}
    \centering
    \includegraphics[width=5in]{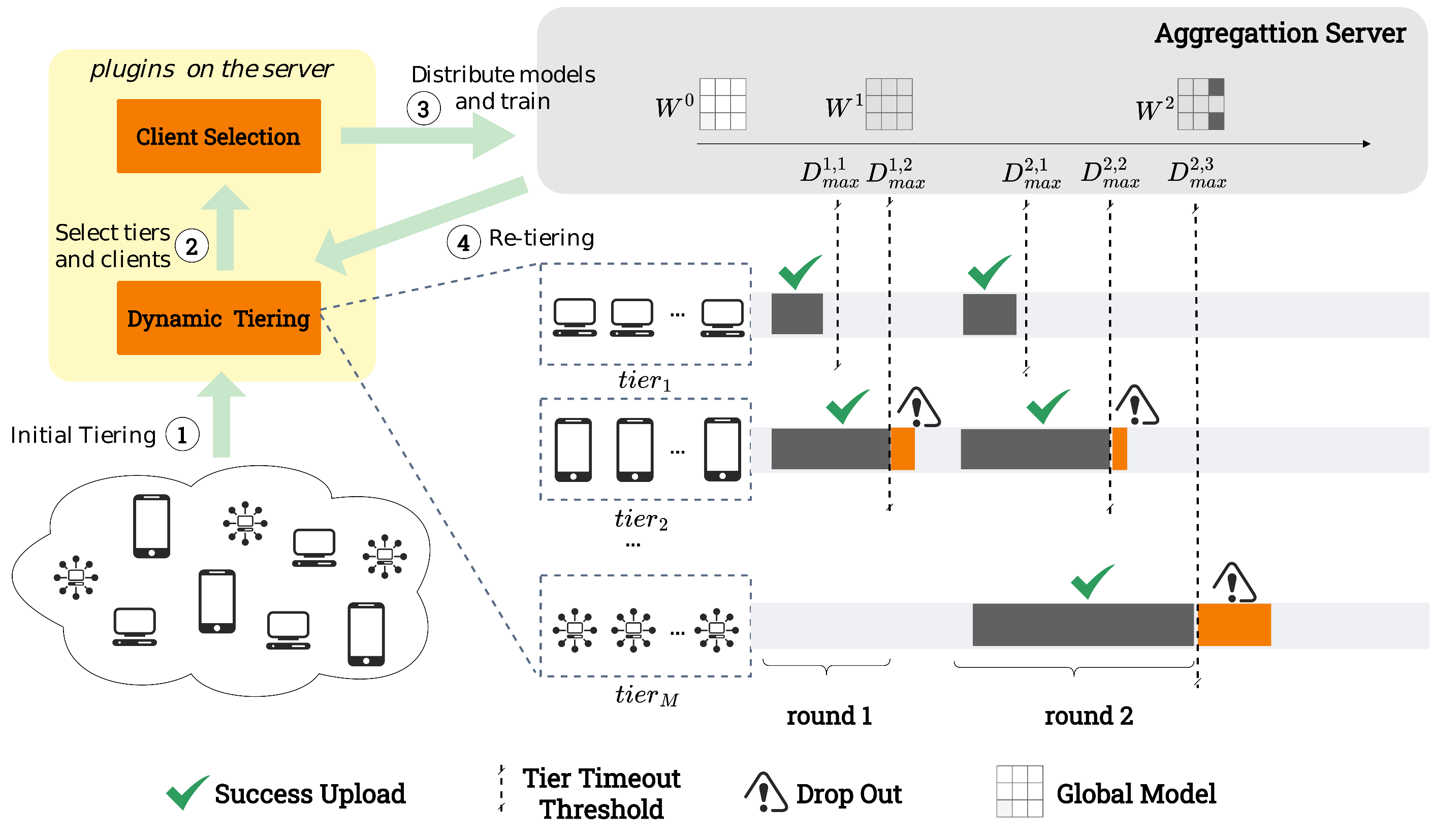}
    \caption{Overview of FedDCT.}
    \label{fig:Overview}
\end{figure}

\subsection{Overview of FedDCT}

FedDCT consists of three main components: 1) Aggregation Server: Responsible for globally synchronizing model updates. 2) Dynamic Tiering Module: Dynamically assesses the response time of clients categorizes them into different tiers, and assigns specific timeout thresholds to each tier. 3) Cross-Tier Client Selection Module: Selects tiers based on the current accuracy changes in the global model, then selects participating devices within each tier based on their training information. The proposed dynamic tiering module and cross-tier client selection module can operate as independent plugins running on the aggregation server. Taking the participation of devices selected as $\{tier_1, tier_2\}$ in the first round and $\{tier_1, tier_2, tier_3\}$ in the second round as an example, the process is illustrated in Fig. \ref{fig:Overview}.
\begin{enumerate}[label=\circled{\arabic*}]
    \item During the initialization phase, the dynamic tiering module evaluates the average response time $t_a$ of all participating devices. Subsequently, clients are stratified into $M$ tiers denoted as $\{tier_1, ..., tier_M\}$ based on the response times of each device. Here, $tier_1$ represents the fastest tier, and $tier_M$ represents the slowest tier. As part of the tiering process, distinct timeout thresholds are assigned to devices within each tier.(\$\ref{dy_tiering})
    \item The client selection module, based on the accuracy change in the globally aggregated model from the previous round, chooses the tier $j$ for participation in the current training round. Subsequently, devices are selected from the tier set $\{tier_1, ..., tier_j\}$ with weighted consideration, forming the set of participating devices $C_r$.(\$\ref{ctcs})
    \item The aggregation server distributes the latest global model $W$ to the selected participating devices. Devices then train their models based on the global model and local data, subsequently returning their training results. For devices that exceed the timeout threshold $D^{j}_{max}$, the server no longer waits for their uploads, marking them as dropout devices. The system undergoes a reevaluation and tiering process for these devices.
    \item The dynamic tiering module updates the average response time based on the actual time usage of all devices in the current training round and subsequently performs a re-tiering process. Unlike approaches such as TiFL \cite{chai2020tifl} and FedAT \cite{chai2021fedat}, which assess devices only in the initialization phase, this dynamic evaluation more accurately reflects the variability in resource heterogeneity within wireless networks.(\$\ref{dy_tiering})
\end{enumerate}

The iterative process of steps \circled{2}-\circled{4} continues until a specified number of training rounds is completed or the model converges to the desired accuracy requirement.

\subsection{Dynamic Tiering}
\label{dy_tiering}

\begin{figure}[h!]
    \centering
    \includegraphics[width=3in]{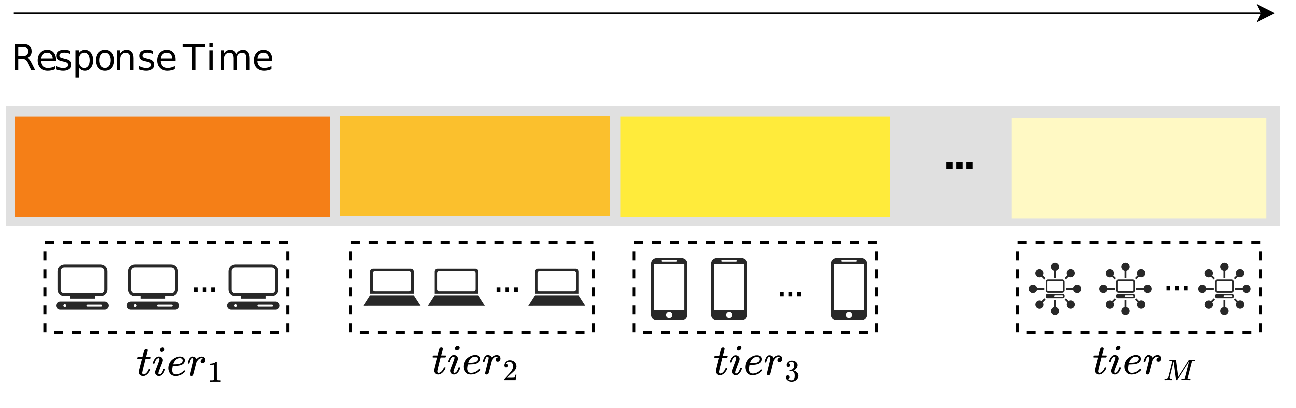}
    \caption{Response time of devices in different tiers. }
    \label{fig:TieringWithTimeout}
\end{figure}

The dynamic tiering module primarily incorporates three main functionalities: 1) Evaluate the average response time of participating devices. 2) Categorize devices into different logical tiers based on their average response times. 3) Calculate the timeout threshold for devices within each tier based on their average response times. Specifically, the module categorizes devices into $M$ tiers based on their average response times $t_a$ during the $ct[i]$ training round. Devices with an average response time $t_a$ exceeding a threshold are considered dropout devices, and they undergo re-evaluation and re-tiering after $\kappa$ rounds.

 \begin{algorithm}[h!]
\caption{Tiering}
\label{alg:Tiering}
\LinesNumbered
\KwIn {the average response time of clients $t_a$, the number of client in tier $Ts$.}
\KwOut {tiering of clients $ts$. }
\BlankLine
\For{client $c$, time $t$ in $t_a$}{
    $tmp[c] = (c,t)$ \;   
}
$tmp = SortAscByTime(tmp)$ \;
\For{index $i$, client $c$ in $tmp$}{
    $ts[i/Ts][i\%Ts] = c$ \;  
}
\textbf{return} $ts$;
\end{algorithm}

In Algorithm \ref{alg:Tiering}, we provide a detailed description of how the dynamic tiering module categorizes devices into $M$ different logical tiers based on their average response time $t_a$. The logical tiers, arranged from low to high, reflect an increasing order of average response times of the devices within each tier. The tiering effect is illustrated in Fig. \ref{fig:TieringWithTimeout}, where devices in tiers $tier_1$ through $tier_M$ exhibit increasing response times, and devices within each tier have approximately similar response times.

\begin{figure}[h!]
    \centering
    \includegraphics[width=3in]{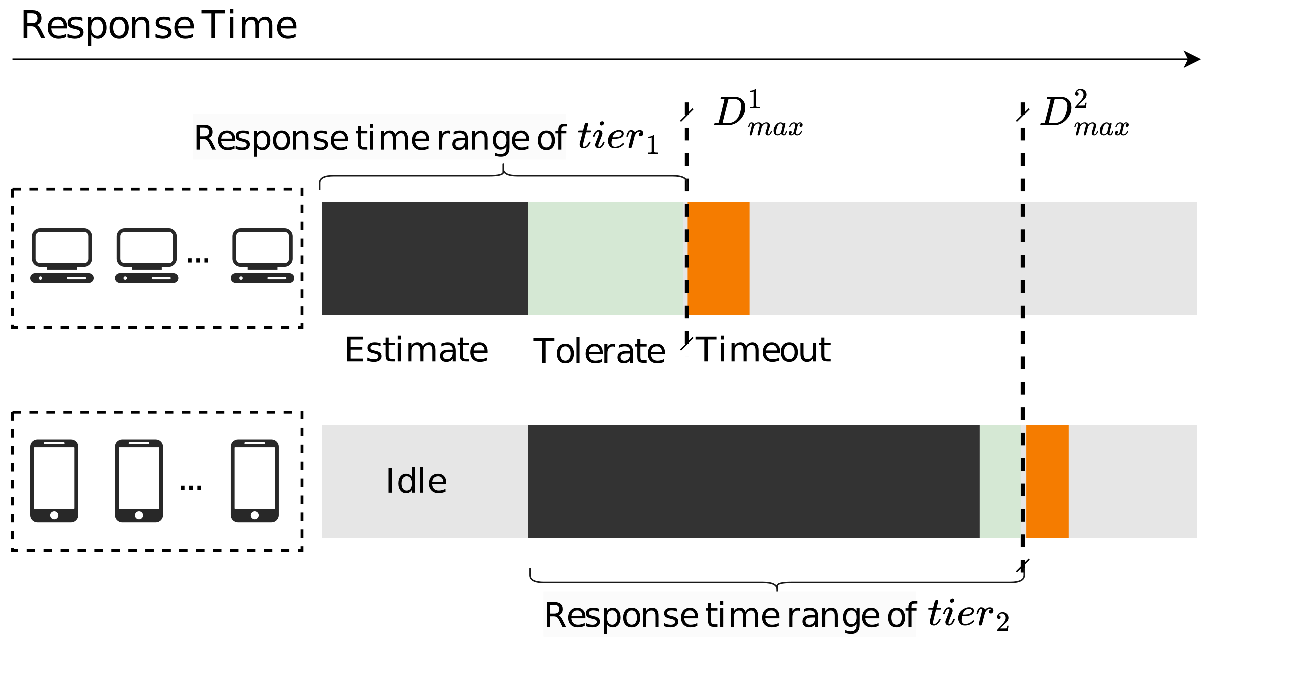}
    \caption{Response time analysis of $tier_1$ and $tier_2$ in a task.}
    \label{fig:Tiering}
\end{figure}

The purpose of setting the timeout threshold is to prevent excessive waiting time caused by resource heterogeneity and dropout devices. However, unlike conventional FL approaches, FL with tiering should adopt more refined timeout thresholds. Therefore, we utilize the average response time of devices in tier $j$, denoted as $\frac{\sum_{i \in ts[t] }t_a[i]}{len(ts[t])}$, multiplied by a tolerance limit $\beta$ as the timeout threshold $D^{j}_{max}$ for that tier. The tolerance limit $\beta$ reflects the degree of tolerance for delayed responses from devices in wireless networks. A larger $\beta$ not only signifies more tolerance for delays, as illustrated in Fig. \ref{fig:Tiering}, but also allows devices that exceed $D^{j}_{max}$ to be deemed as dropout devices. These devices undergo $\kappa$ rounds of re-evaluation until normal completion of $\kappa$ rounds, after which they are re-tiered and reintroduced into subsequent training. At the same time, we also set a maximum timeout threshold of $\Omega$ to limit the average training time of this tier to be too long.

\begin{equation}
   D^{j}_{max} = min( \frac{\sum_{i \in ts[t] }t_a[i]}{len(ts[t])} \times \beta , \Omega )
\end{equation}

\begin{algorithm}[h!]
\caption{Client Selection}
\label{alg:client_selection}
\LinesNumbered
\KwIn {current tier $j$, last test accuracy $\upsilon_{last}$, global model $W$, tiering of clients $ts$, the number of client training $ct$, the number of client selection in a tier $\tau$.}
\KwOut {the clients selection $C_r$. }
\BlankLine
$\upsilon = Evaluation(W, TestData)$ \;
\If{$\upsilon \geq \upsilon_{last}$}{
    $j = max(j - 1, 1)$ \;
}\Else{
    $j = min(j + 1, T)$ \;
}

\For{tier $t=1$ $\to$ $j$}{
     \For{client $c$ in $ts[t]$}{
        $probs[c]= \frac{1/ct[c]}{\sum_{i\in ts[t]}1/ct[i]}$ \;
    } 
    $clients$  = (select $\tau$ clients from tier $t$ with $probs$) \;
    $C_r$ $\gets$ $clients$ \;
}
\textbf{return} $C_r$;
\end{algorithm}

\subsection{Cross-Tier Client Selection} 
\label{ctcs}

The cross-tier client selection module selects participating devices for FL to achieve fast response performance while ensuring model convergence.
This module is divided into two main steps: 1) Selecting participating tiers and 2) Selecting devices within those tiers.

Initially, based on the tiering characteristics described in \$\ref{dy_tiering}, the expectation is to select tiers from low to high. If devices in the fast-responding $tier_{j}$ contribute to the convergence of the global model, devices from $tier_{j+1}$ are not selected. The change in accuracy of the global model $\upsilon$ is used as a criterion. If the currently evaluated accuracy $\upsilon$ after aggregation is higher than the accuracy $\upsilon_{last}$ in the previous round, it indicates that the devices from $tier_j$ currently used can still contribute to the convergence of the global model. To reduce training time, an attempt can be made to select devices from $tier_{j-1}$ in the next round. It's important to note that, to minimize idle waiting time for devices, as illustrated in Fig. \ref{fig:Tiering}, the proposed approach allows for cross-tier selection. In other words, when selecting $tier_{j}$, devices from $\{tier_{1},...,tier_{j}\}$ are actually chosen.

\begin{equation}
    j = \begin{cases}
    min(j + 1, M), & \upsilon < \upsilon_{last} \\
    max(j - 1, 1), & \upsilon \geq \upsilon_{last} \\
    \end{cases}
\end{equation}

To prevent significant differences in the participation frequency among devices within tiers, which may lead to model drift \cite{huang2020efficiency}, we increase the selection probability of devices with fewer participation times when selecting nodes within tiers. Therefore, we allocate different selection probabilities $probs$ based on the participation times $ct$ of devices in $tier_{j}$. Finally, according to the selection probabilities $probs$ of devices within tiers, $\tau$ devices $C_r$ are chosen to participate in training for this round, as depicted in Algorithm \ref{alg:client_selection}.

\section{Experimental Evaluation}\label{result}

We referred to a portion of the implementation methods from Fedlab\cite{smile2021fedlab} and implemented FedDCT and other FL baseline methods using PyTorch. All experiments were conducted on a high-performance server with 2 $\times$ Intel(R) Xeon(R) Gold 6230 CPUs, 128GB of memory, and 2 $\times$ NVIDIA Tesla V100 FHHL GPUs. We simulated a scenario where one server and 50 clients participated in FL training on this machine.

\subsection{Experimental Setup}
We conducted experiments on three commonly used datasets, MNIST\cite{lecun1998gradient}, CIFAR-10\cite{cifcar10}, and Fashion-MNIST\cite{xiao2017fashion}. Two classic neural network models, CNN and ResNet8, were employed for training. We used the CNN model for training on MNIST and Fashion-MNIST datasets and the ResNet8 model following the approach in the literature \cite{ijcai2022p308} for training on CIFAR-10. The proposed approach will be compared with three classic algorithms for synchronous (FedAvg\cite{mcmahan2017communication}), asynchronous (FedAsync\cite{xie2019asynchronous}), and tiered FL (TiFL\cite{chai2020tifl}).

We used momentum as the optimization algorithm with a learning rate of 0.001 and momentum of 0.9. For each dataset, we trained with the following configurations: local epoch = 1, batch size = 10, $\tau$ = 5, $\beta$ = 0.1, $\Omega$ = 30s, $\kappa = 3$. We used the same parameters for other FL approaches. The default number of selected clients for training in each round was 5, but for FedDCT, the number of selected clients per round varied with the selected tier.

To simulate the response time differences caused by resource heterogeneity in wireless networks, we assigned random response delays with a variance of $2$ from a Gaussian distribution with expectations of $\{5, 10, 15, 20, 25\}$ seconds for devices. Additionally, to simulate dropout occurrences, we randomly added delays in the range of $(30-60)$ seconds during training, controlled by the dropout rate $\mu$ to determine the probability of its occurrence. Finally, to analyze the training effects under different data distribution scenarios, we randomly assigned a main class to each client, where \#\% of the data in that device belonged to the main class, and the remaining data belonged to the other classes.

 \begin{table}[h!]
    \centering
     \caption{Comparison of the best average accuracy and time which reach the preset accuracy of each baseline algorithm. \# represents the percentage of primary class label in each client. Accuracy shows the best average accuracy achieved after convergence. Time represents the time taken by the model to converge to the specified precision(s). For CIFAR-10, Fashion-MNIST, and MNIST, the convergence accuracy is preset as 0.7, 0.88, and 0.98, respectively (CIFAR-10 \#=0.7 is preset as 0.6 separately).  impr.(a) and (b) represent the improved training accuracy of FedDCT and the reduced time of convergence to the specified accuracy compared with the best baseline FL method, respectively.}
    \label{table:result}%
    \begin{tabular}{c|ccccccc}
        \toprule 
        \multicolumn{2}{c}{Dataset} & \multicolumn{4}{c}{CIFAR-10} & \multicolumn{1}{c}{Fashion-MNIST} & \multicolumn{1}{c}{MNIST} \\
        \multicolumn{2}{c}{(\#Non-IID)} & IID & \#0.3 & \#0.5 & \#0.7 & \#0.7 & \#0.7\\
        \midrule 
        \multirow{2}{*}{FedAvg} & Accuracy & 0.7843 & 0.7407 & 0.7150 & 0.6592 &0.8914 & 0.9892 \\
         & Time(s) & 1617.0 & 2403.5 & 3416.2 & 3033.8 & 2544.1 & 1481.9 \\
        \midrule 
        \multirow{2}{*}{TiFL} & Accuracy & 0.7826 & 0.7401 & 0.7071 & 0.6475 & 0.8862 & 0.9894 \\
          & Time(s) & 1980.8 & 1945.5 & 3389.9 & 2363.2 & 2431.4 & 1261.6 \\
        \midrule 
        \multirow{2}{*}{FedAsync} & Accuracy  & 0.7718 & 0.7252 & 0.7001 & 0.6234 & 0.8786 & 0.9868 \\
         & Time(s)  & 3709.6 & 4885.5 & 6268.6 & 7435.5 & 6417.0 & 2427.4 \\
        \midrule 
        \multirow{3}{*}{FedDCT}  & Accuracy& \textbf{0.7920} & \textbf{0.7526} & \textbf{0.7287} & \textbf{0.6897} & \textbf{0.9080} & \textbf{0.9897} \\
          & Time(s) & \textbf{685.6} & \textbf{618.5} & \textbf{1479.4} & \textbf{1077.3} & \textbf{965.8} & \textbf{864.7} \\
          & impr.(a) & 0.98\% & 1.60\% & 1.91\% & 4.62\% & 1.86\% & 0.03\% 
          \\
           & impr.(b) & 57.6\% & 68.2\% & 56.3\% & 54.4\% & 60.2\% & 31.4\%
          \\
        \bottomrule 
    \end{tabular} 
\end{table}

\subsection{ Experimental Results}

Table \ref{table:result} presents the best average accuracy and the time spent to reach the preset accuracy for all datasets. The results show that, across all six scenarios, the proposed approach achieved an average accuracy improvement of 1.83\% and reduced time overhead by 54.7\% compared to the optimal baseline. Under the same experimental configuration, FedDCT consistently achieved higher convergence accuracy and significantly reduced convergence time in all experiments. Particularly, the improvement compared to TiFL indicates that 1) the dynamic tiering in the proposed approach is more accurate and adaptable to changes in the dynamic environment, and 2) the selection of devices across tiers effectively exploits device performance, enhancing convergence speed. Meanwhile, we observed that TiFL does not perform well in the presence of unexpected dropouts, leading to suboptimal convergence accuracy and time.

\begin{figure}[h!]
\centering
\subfigure[$\#$=0.3]{
    \begin{minipage}[t]{0.3\linewidth}
        \centering
        \includegraphics[width=\linewidth]{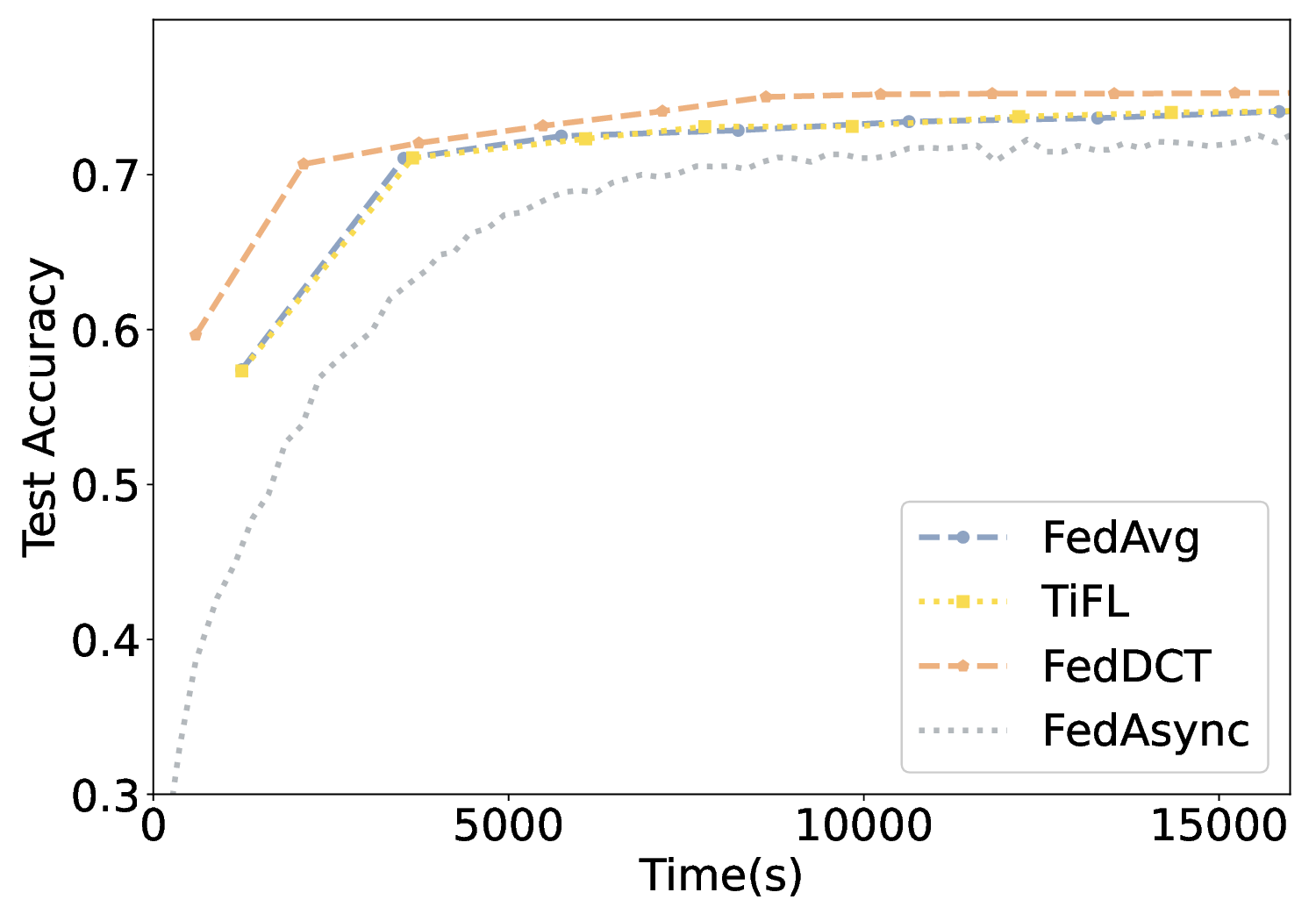}\\
        \includegraphics[width=\linewidth]{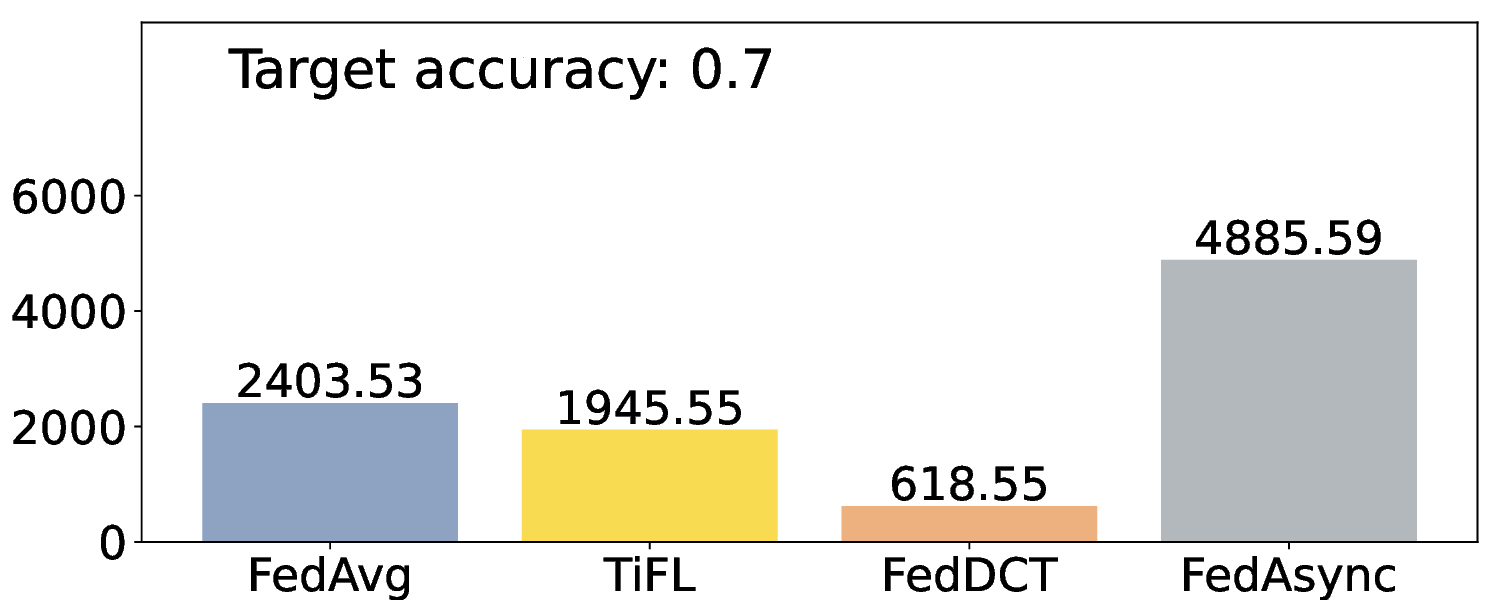}\\
    \end{minipage}%
}%
\subfigure[$\#$=0.5]{
    \begin{minipage}[t]{0.3\linewidth}
        \centering
        \includegraphics[width=\linewidth]{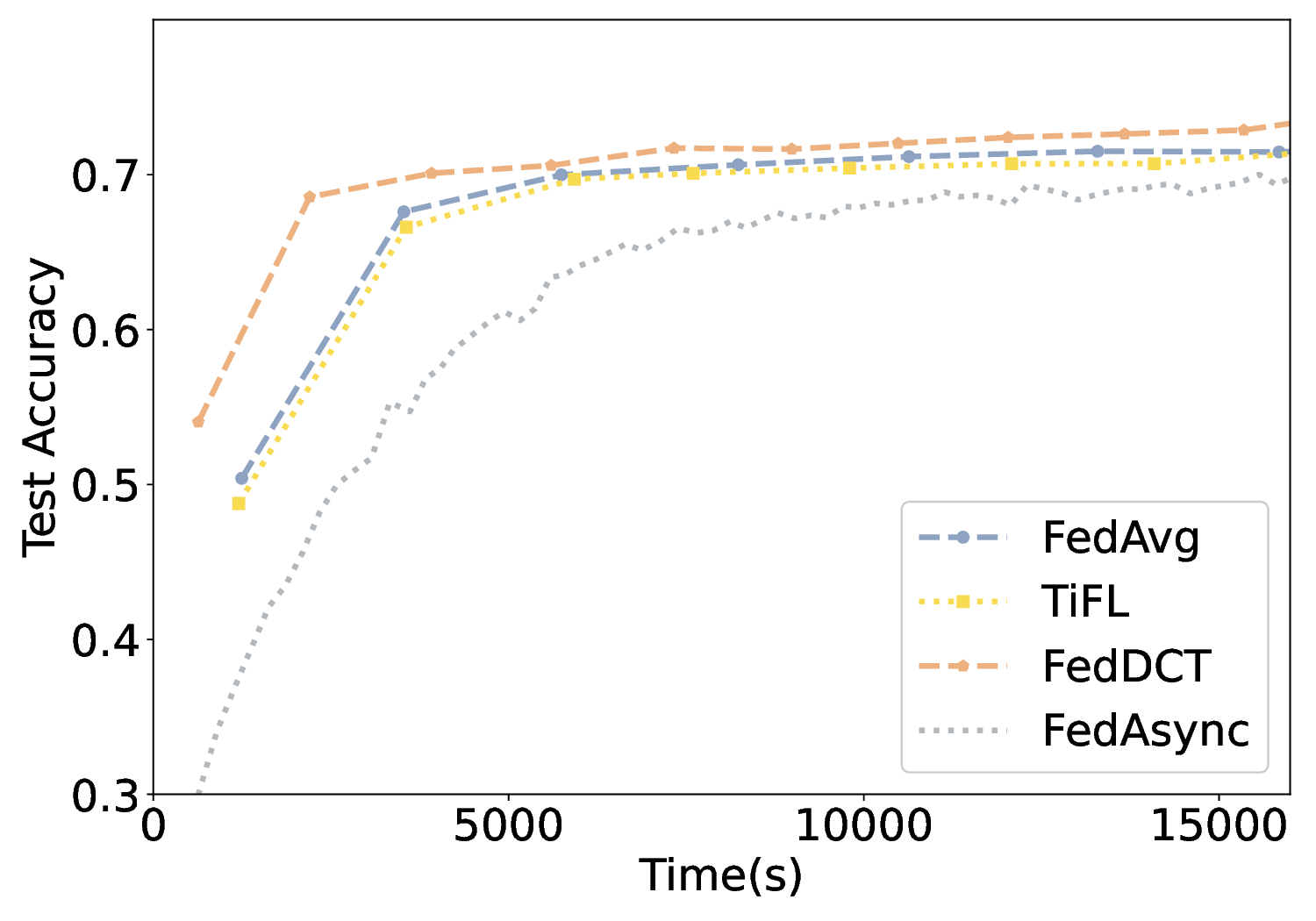}\\
        \includegraphics[width=\linewidth]{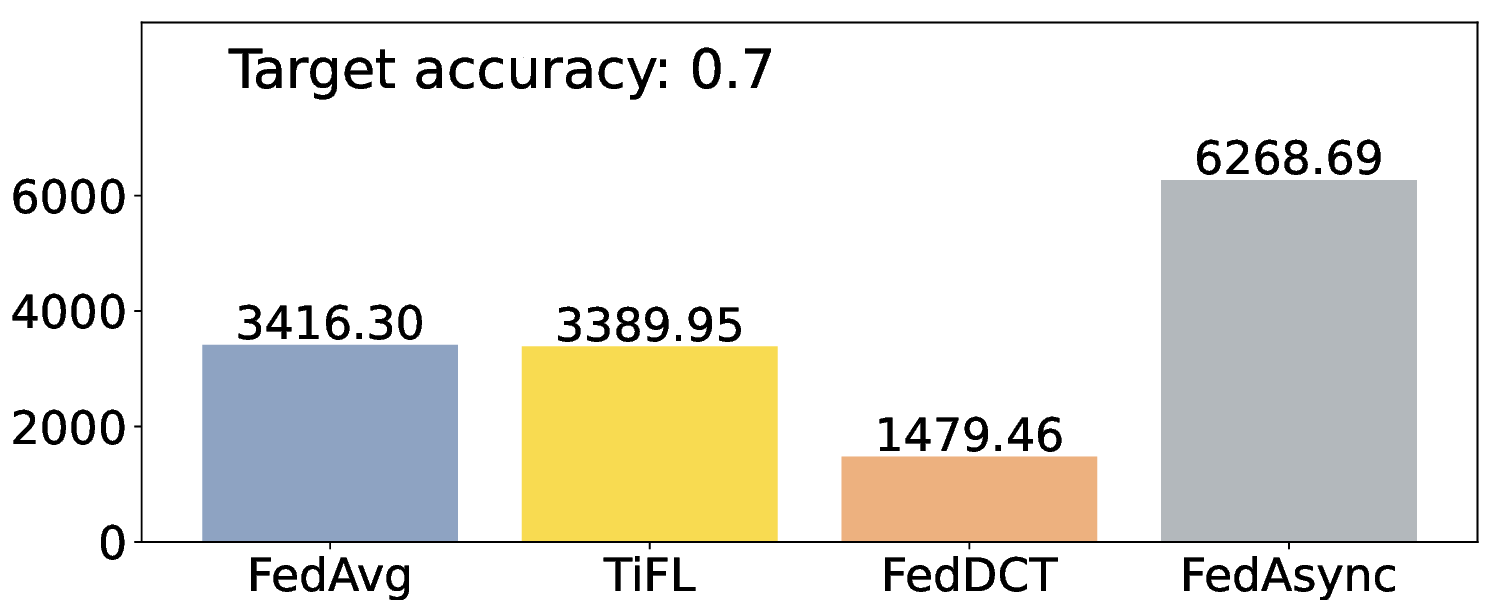}\\
    \end{minipage}%
}%
    \subfigure[$\#$=0.7]{
    \begin{minipage}[t]{0.3\linewidth}
        \centering
        \includegraphics[width=\linewidth]{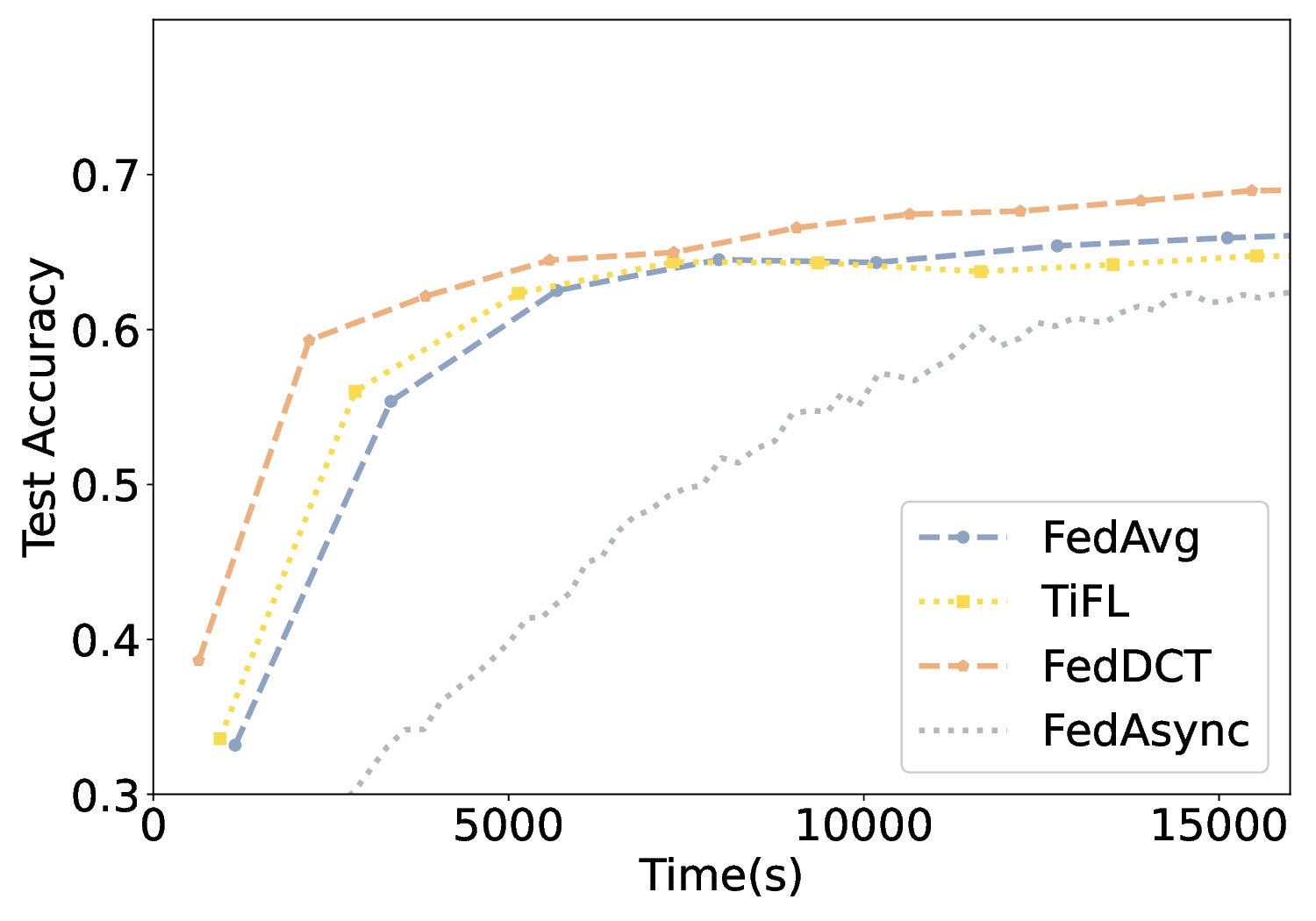}\\
        \includegraphics[width=\linewidth]{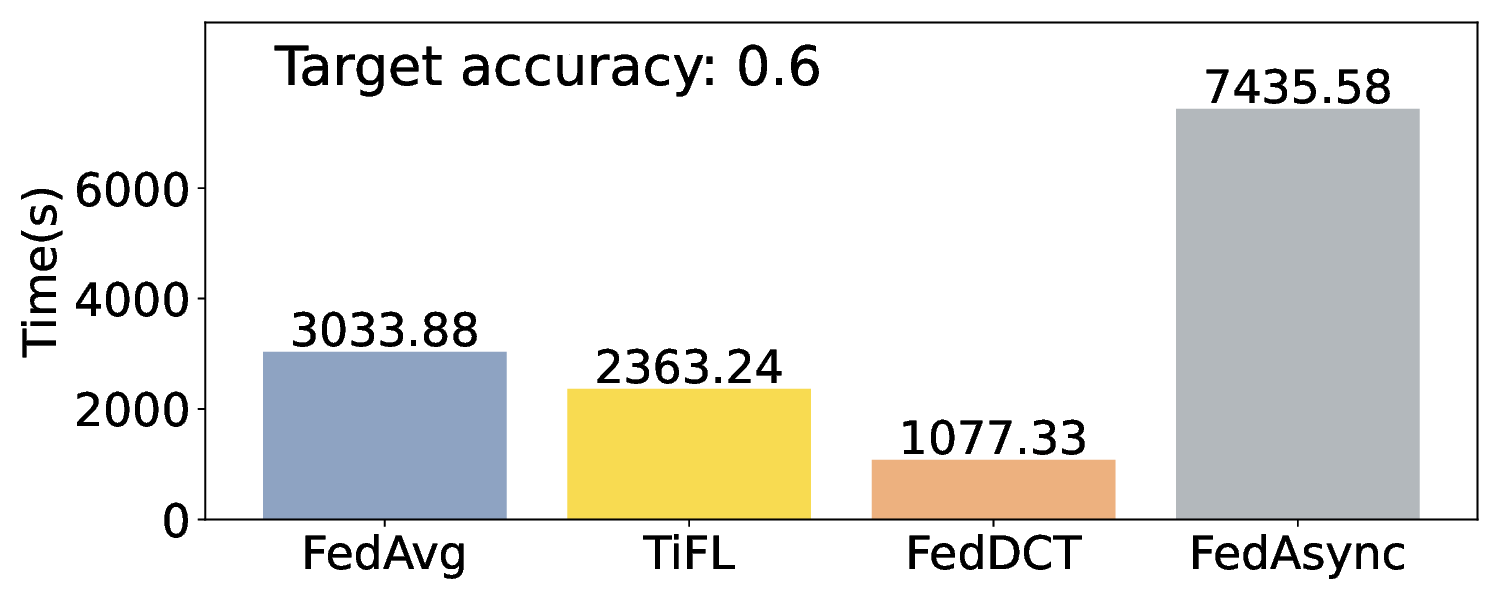}\\
    \end{minipage}%
}%
\centering
\caption{The effect of different $\#$ on training.}
\label{noniid}
\end{figure}

\begin{figure}
\centering
\subfigure[$\mu$=0]{
    \begin{minipage}[t]{0.3\linewidth}
        \centering
        \includegraphics[width=\linewidth]{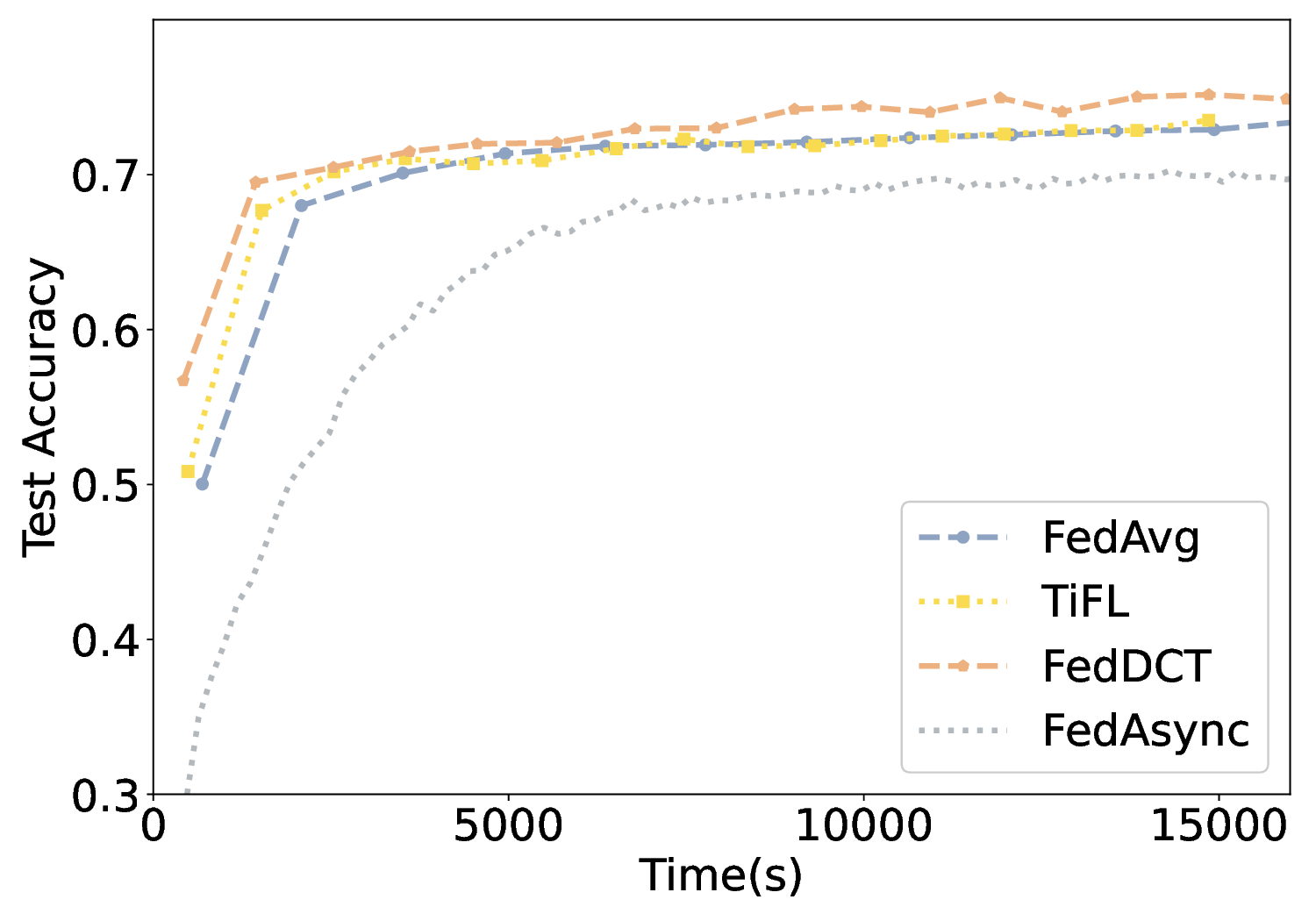}\\
        \includegraphics[width=\linewidth]{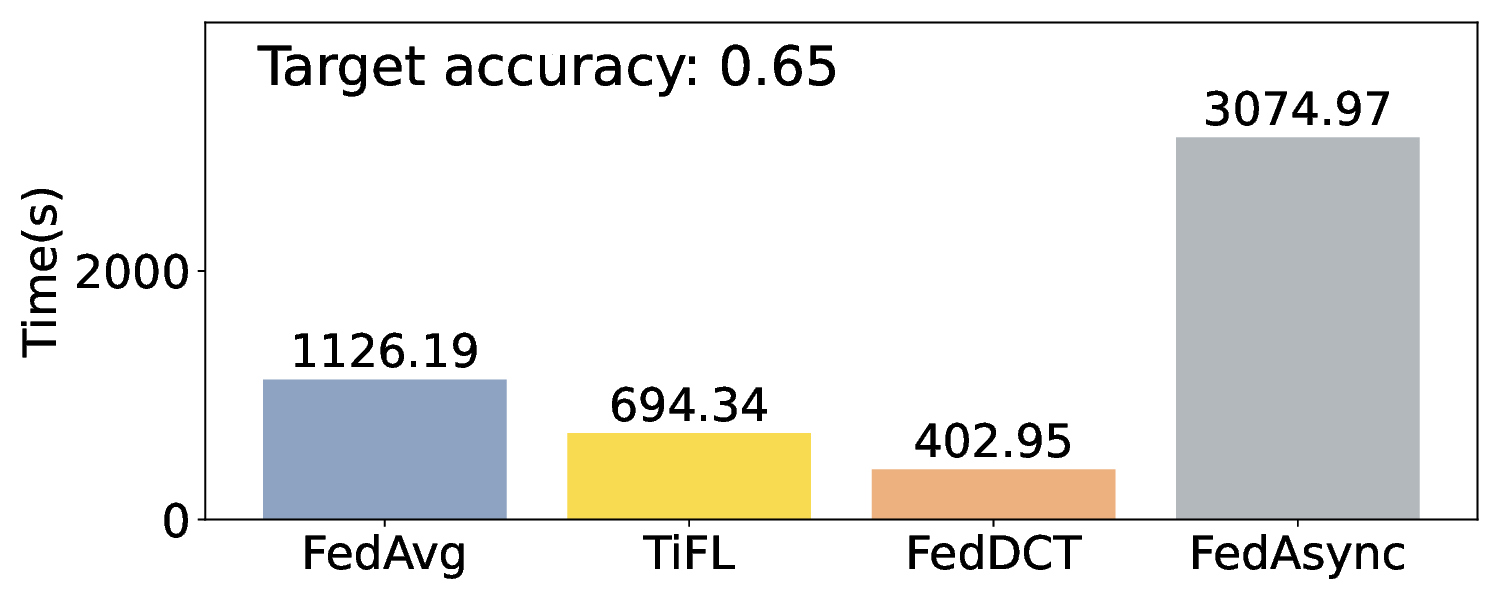}\\
    \end{minipage}%
}%
\subfigure[$\mu$=0.1]{
    \begin{minipage}[t]{0.3\linewidth}
        \centering
        \includegraphics[width=\linewidth]{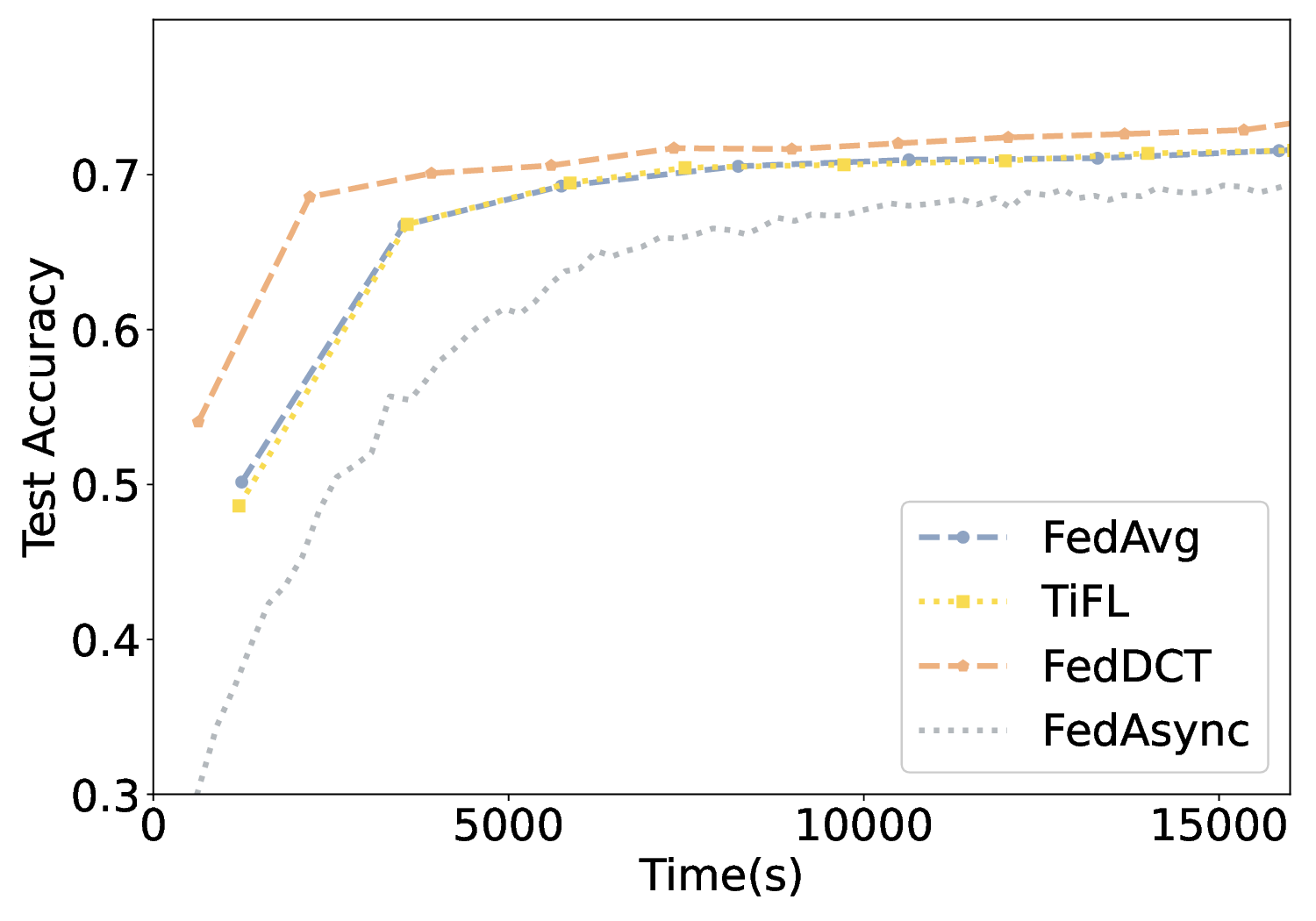}\\
        \includegraphics[width=\linewidth]{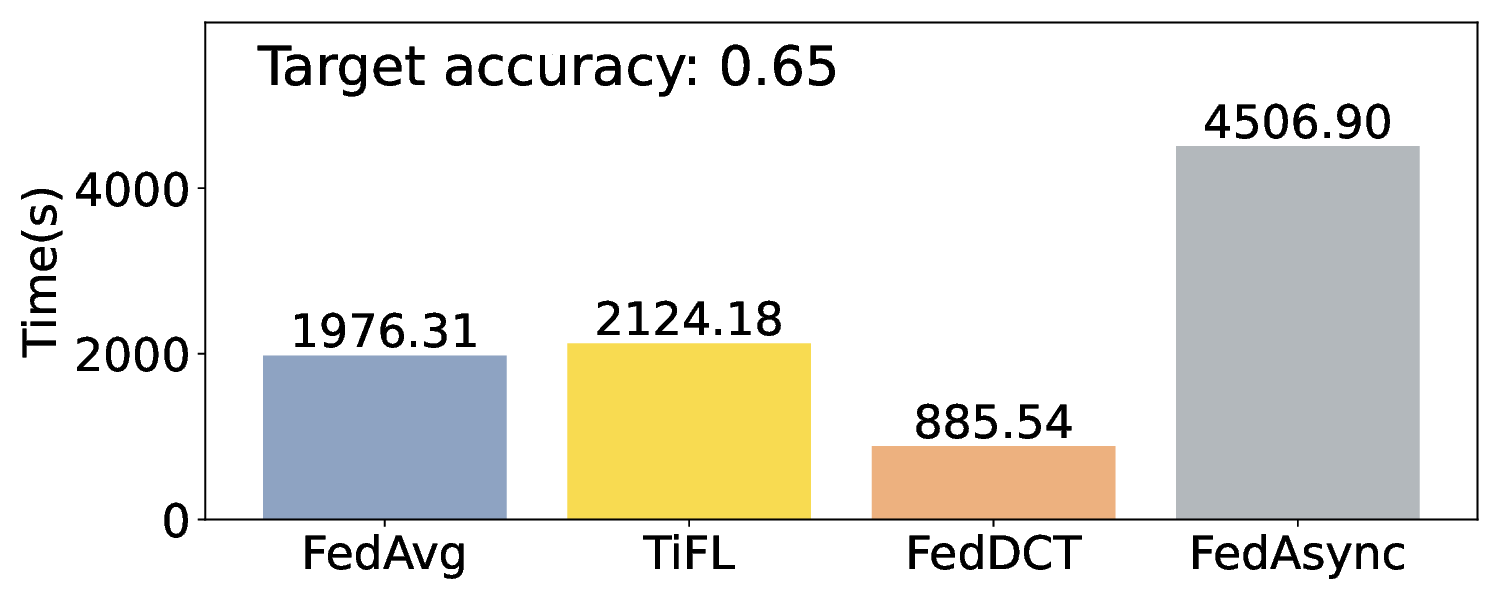}\\
    \end{minipage}%
}%
    \subfigure[$\mu$=0.4]{
    \begin{minipage}[t]{0.3\linewidth}
        \centering
        \includegraphics[width=\linewidth]{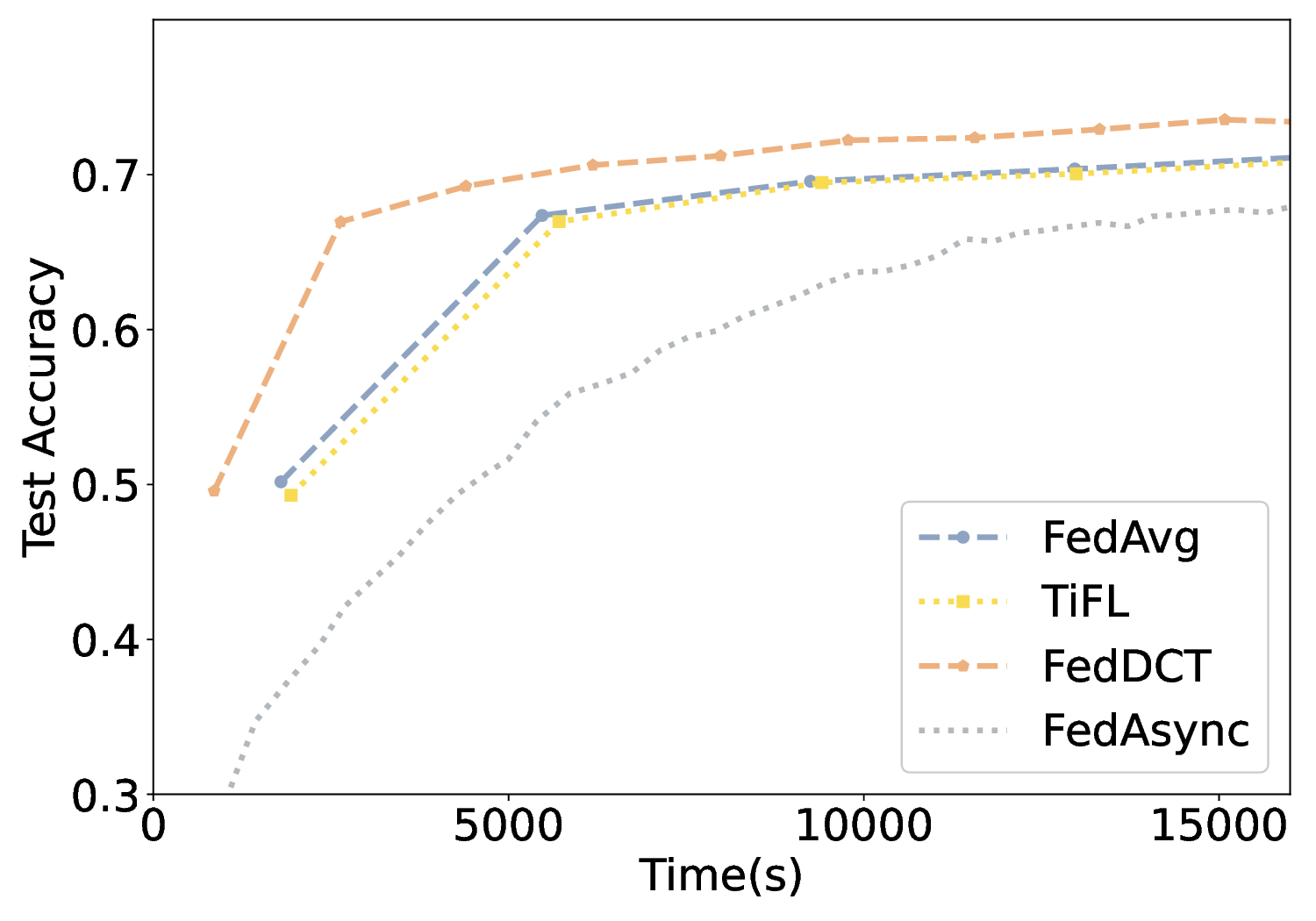}\\
        \includegraphics[width=\linewidth]{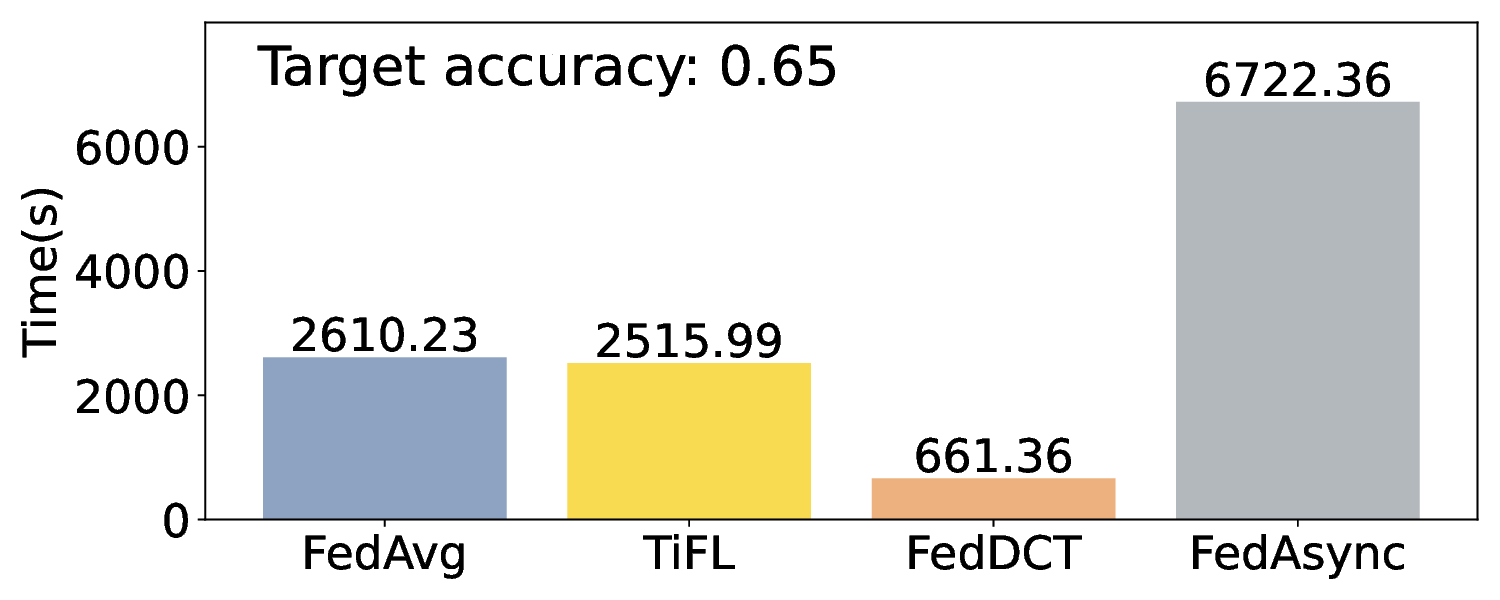}\\
    \end{minipage}%
}%
\centering
\caption{The effect of different $\mu$ on training.}
\label{drop}
\end{figure}

Fig. \ref{noniid}-\ref{drop} illustrates the training performance of all schemes under different data distributions $\#$ and various dropout rates $\mu$. Fig. \ref{noniid} indicates that the proposed scheme performs well under different data distributions. Although the overall convergence accuracy decreases with the increasing heterogeneity of data distribution, our scheme can still achieve faster convergence and higher final convergence accuracy compared to other baseline schemes. Fig. \ref{drop} demonstrates that as the dropout rate $\mu$ increases, the overall convergence time also gradually increases. However, we observe that the impact of the dropout rate $\mu$ on the convergence of FedDCT is not significant. This is attributed to the dynamic tiering module in FedDCT, which can significantly alleviate the impact of device dropouts on FL.

\begin{figure}[h!]
    \centering
    \subfigure[Stable Network]{
        \begin{minipage}[t]{0.45\linewidth}
            \centering
            \includegraphics[width=\linewidth]{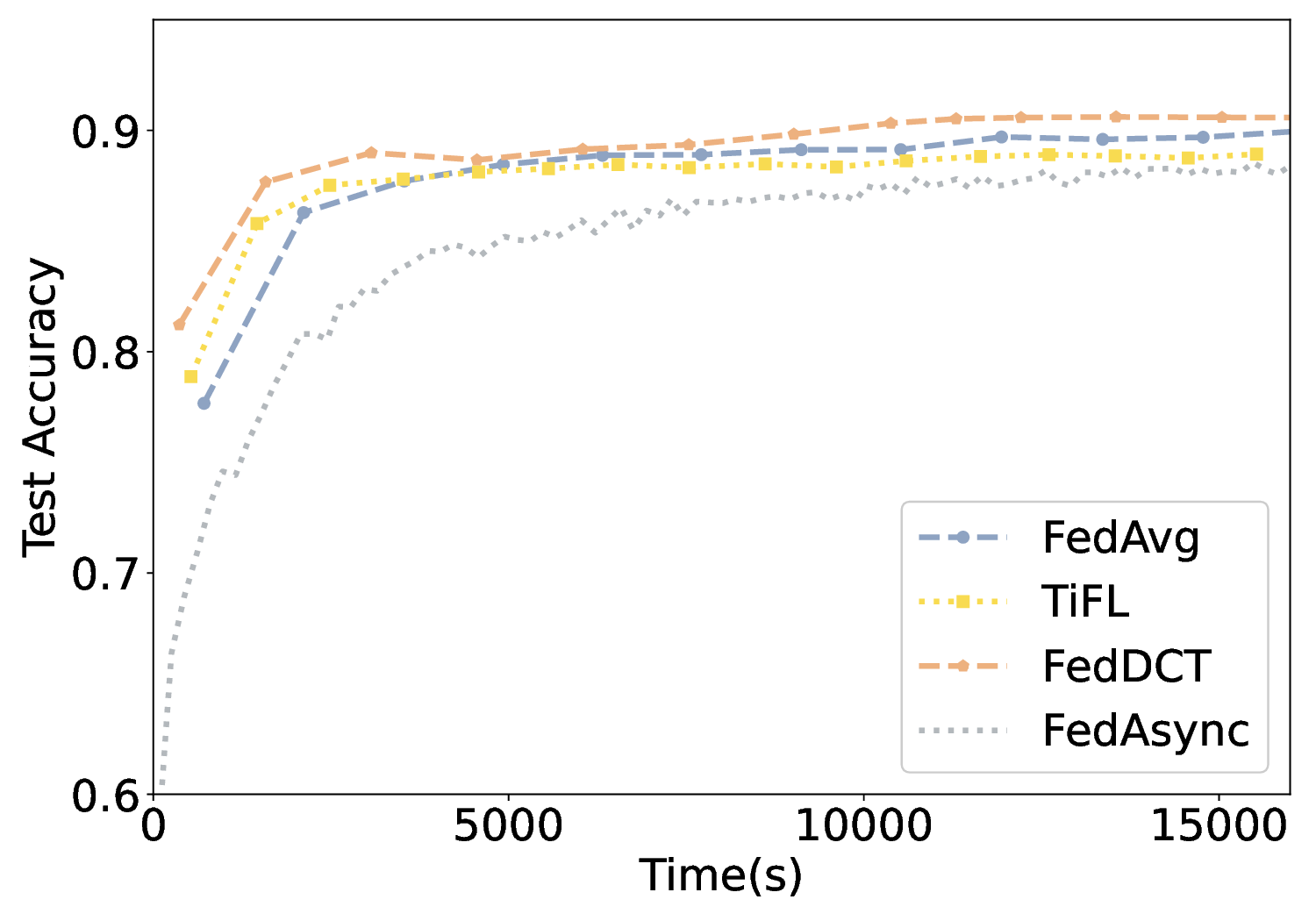}\\
            \vspace{0.02cm}
            \includegraphics[width=\linewidth]{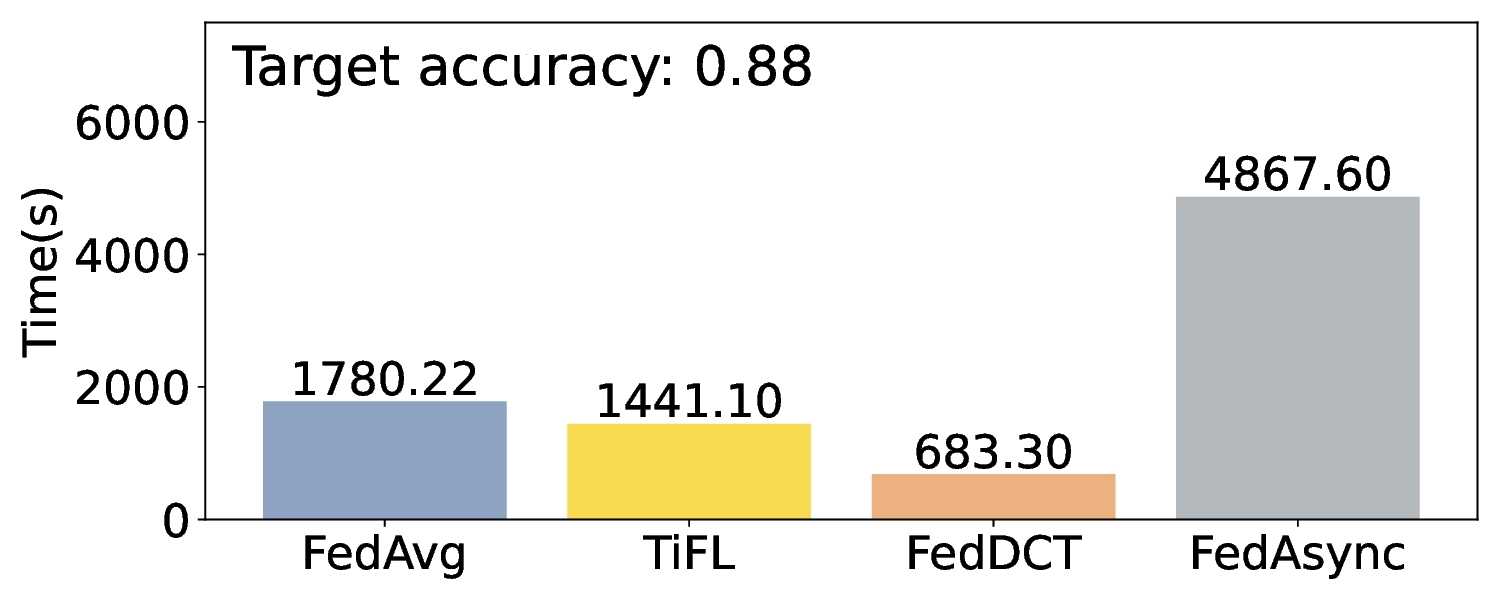}\\
            \vspace{0.02cm}
        \end{minipage}%
    }%
    \subfigure[Complex Network]{
        \begin{minipage}[t]{0.45\linewidth}
            \centering
            \includegraphics[width=\linewidth]{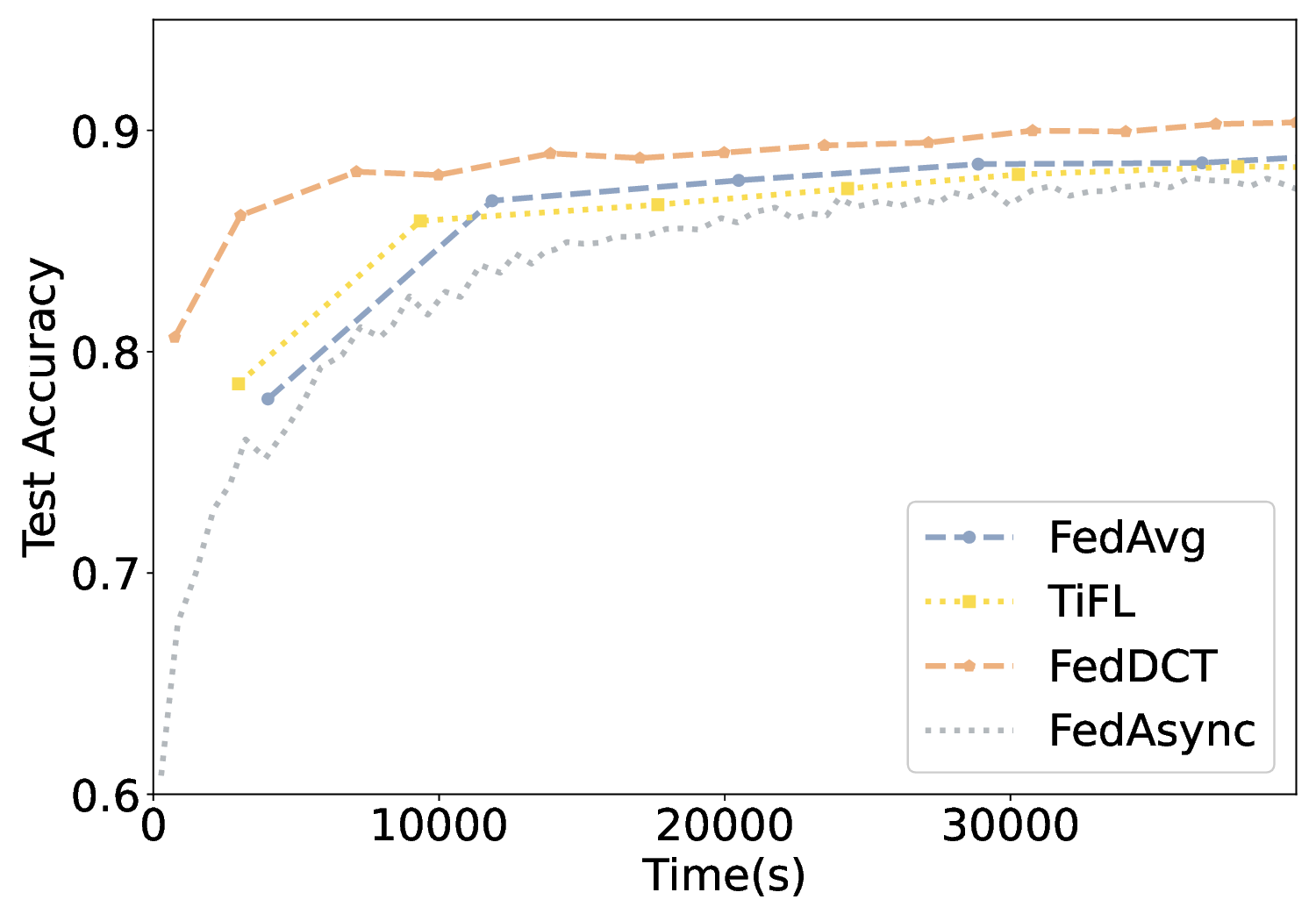}\\
            \vspace{0.02cm}
            \includegraphics[width=\linewidth]{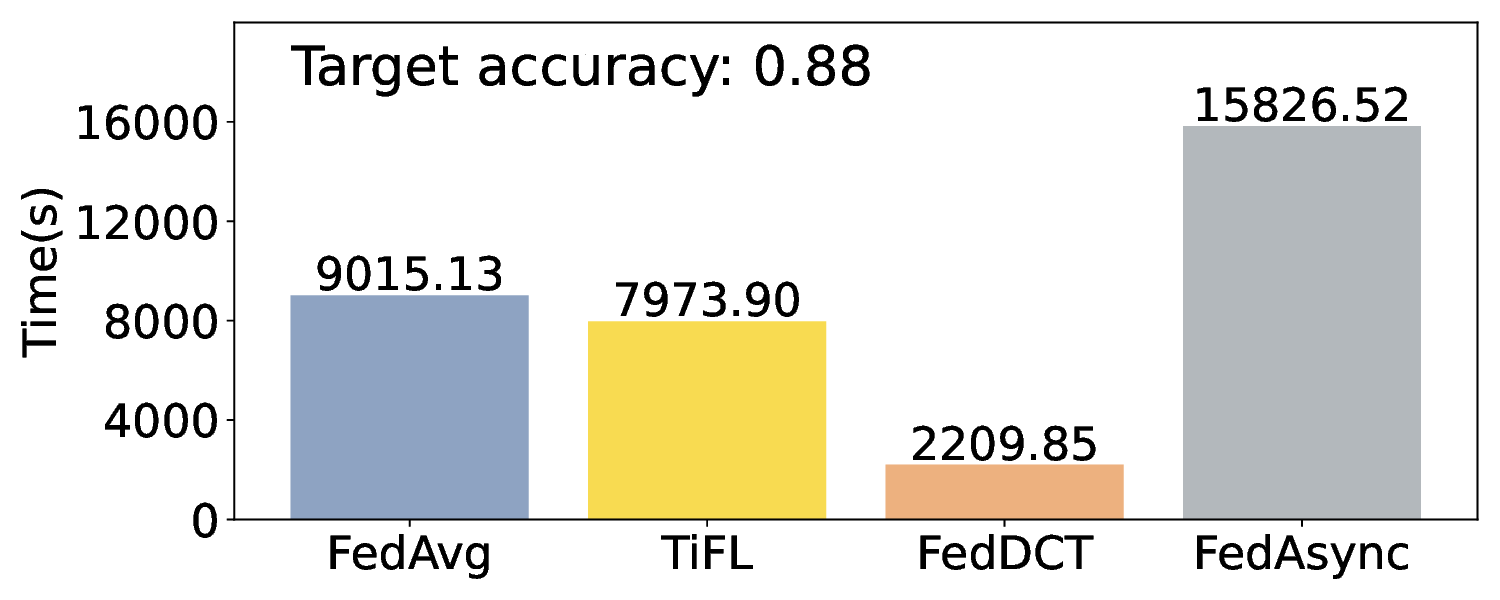}\\
            \vspace{0.02cm}
        \end{minipage}%
    }%
    \centering
    \caption{Training performance under different network environments.}
    \label{diff_env}
\end{figure}

Fig. \ref{diff_env} presents the training performance of all schemes under different network environments. Specifically, in Fig. \ref{diff_env}(a), we set the dropout rate $\mu$ to 0, and in (b), we intensify the response time differences among devices, with response time expectations set to $\{1, 3, 10, 30, 100\}$ seconds. The results indicate that the proposed scheme exhibits good robustness, achieving favorable results in various network environments.

\begin{figure}[h!]
    \centering
    \includegraphics[width=3in]{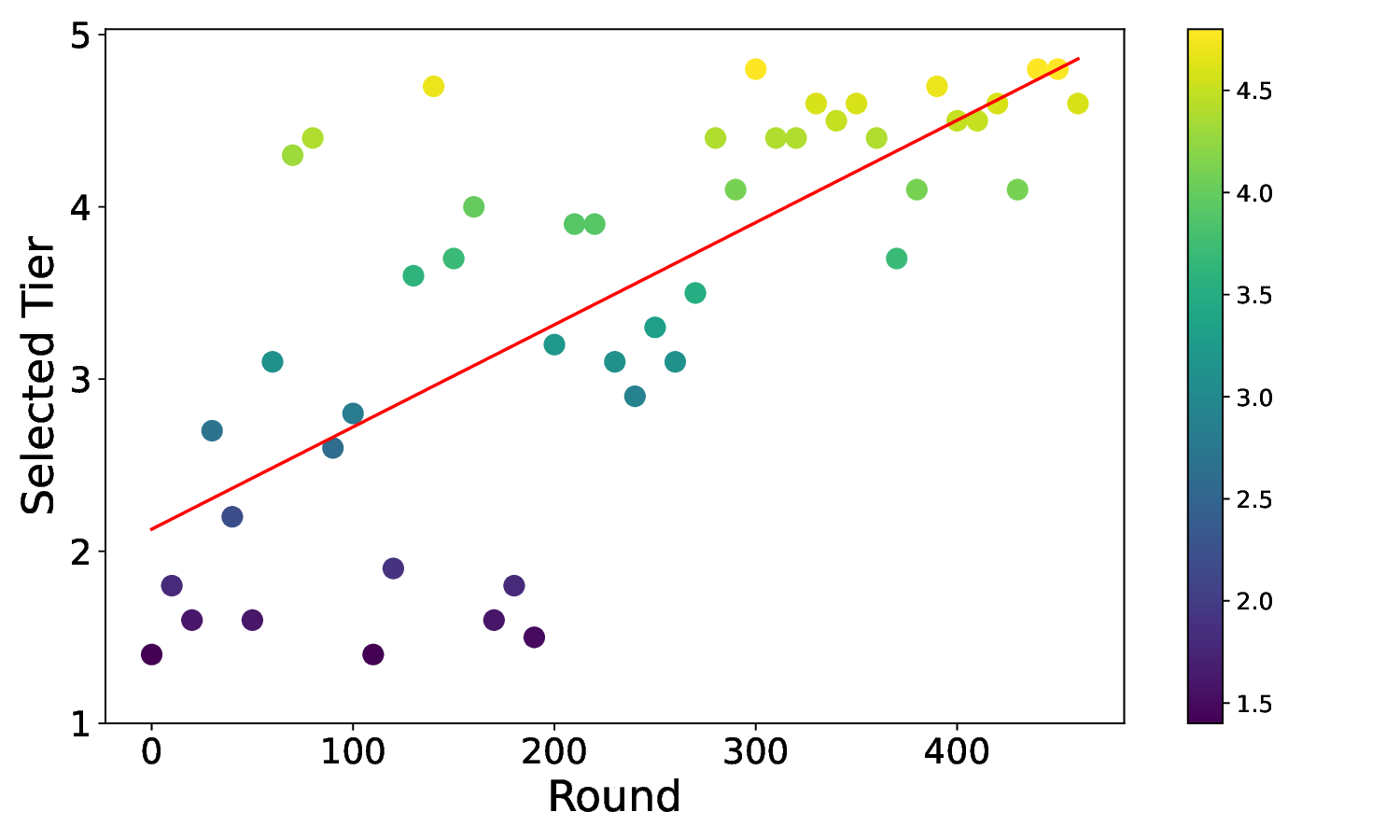}\\
    \centering
    \caption{The changes of the selected tier during the training.}
    \label{tier_choice}
\end{figure}

Finally, to explore why FedDCT could converge faster, we recorded the selected tier during the training process, averaged it every 10 rounds, and fitted it with a linear regression model. As shown in Fig. \ref{tier_choice}, the overall trend of the selected tier increases with training rounds. It is consistent with the expectations of the proposed design. FedDCT first uses the clients in the tier with a short training time for training until it is difficult to improve the accuracy of the global model, and then uses the clients in the other tier with a longer training time.

\section{Conclusion} \label{conclusion}
To mitigate the adverse impact of wireless networks on the training of FL, this paper proposes a novel dynamic cross-tier federated learning Framework. FedDCT adopts a dynamic tiering approach to reduce waiting times during training caused by resource disparities and unexpected device dropouts, thereby enhancing the efficiency of a single training round. Furthermore, we design a cross-tier client selection algorithm, enabling FedDCT to effectively utilize device training information for device selection, thereby improving overall convergence efficiency and accuracy. Experimental results demonstrate that our approach outperforms traditional solutions in wireless networks, achieving superior convergence accuracy and speed.

\begin{credits}
\subsubsection{\ackname} The research was supported in part by the Guangxi Science and Technology Major Project (No. AA22068070), the National Natural Science Foundation of China (Nos. 62166004,U21A20474), the Basic Ability Enhancement Program for Young and Middle-aged Teachers of Guangxi (No.2022KY0057, 2023KY0062), Innovation Project of Guangxi Graduate Education (Nos. XYCBZ2024025).
\end{credits}

\bibliographystyle{splncs04}
\bibliography{myref}

\end{document}